\documentclass[aps,twocolumn,showpacs,preprintnumbers, floats]{revtex4}
\usepackage{amsfonts}
\usepackage{amsmath}
\usepackage{graphicx}
\usepackage{dcolumn}
\usepackage{bm}

\begin{document}

\preprint{}
\title{Conduction through 71$^o$ domain walls in BiFeO$_3$ thin films}

\author{S. Farokhipoor and B. Noheda}
\email{b.noheda@rug.nl}
\affiliation{Zernike Institute for Advanced Materials, University of Groningen, 9747 AG Groningen, The Netherlands.}

\date{\today}

\begin{abstract}
Local conduction at domains and domains walls is investigated in BiFeO$_3$ thin films containing mostly 71$^o$ domain walls. Measurements at room temperature reveal conduction through 71$^o$ domain walls. Conduction through domains could also be observed at high enough temperatures. It is found that, despite the lower conductivity of the domains, both are governed by the same mechanisms: in the low voltage regime electrons trapped at defect states are temperature-activated but the current is limited by the ferroelectric surface charges; in the large voltage regime, Schottky emission takes place and the role of oxygen vacancies is that of selectively increasing the Fermi energy at the walls and locally reducing the Schottky barrier. This understanding provides the key to engineering conduction paths in oxides.

\end{abstract}



\maketitle

Twin walls in ferroic materials provide highly localized regions of large strain gradients and local symmetry breaking in which the properties of the host material can be largely modified. Ekhard K. H. Salje, a pioneer in research on twin walls, has recently gathered the most prominent examples in the past couple of decades, when twin walls in oxides have been reported to show distinct functional properties, including electrical polarization in non-polar CaTiO$_3$ or superconductivity in insulating WO$_3$.\cite{Sal09}. In addition, with the renewed interest in multiferroic materials  (magnetic ferroelectrics), the enhanced coupling of these two order parameters and the distinct magnetoelectric response at domain walls of multiferroic oxides have been subject of much recent attention\cite{Pri97,Gol03,Mos06,Dar10}. Twin domain walls in antiferromagnetic TbMnO$_3$ have been proposed as the origin of net magnetic moment in TbMnO$_3$ thin films\cite{Dau09}; while the net magnetic moment in twin domain walls in antiferromagnetic BiFeO$_3$ is believed to cause the observed exchange bias in a BiFeO$_3$-permalloy bilayer\cite{Mar08}

\begin{figure}
\includegraphics[width=7cm]{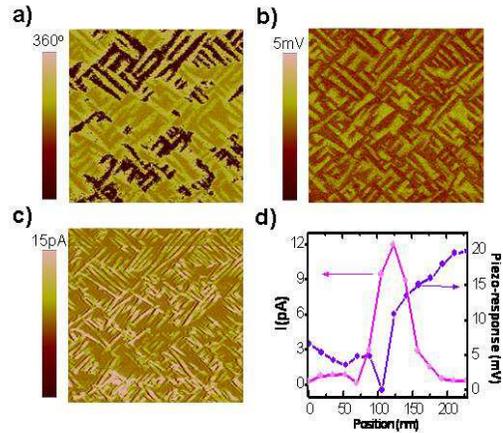}
\caption{(Color online) In-plane piezo-force microscopy images of a 4$\mu $m x4$\mu $m area. Both amplitude (a) and phase (b) images are needed to determine the type of domain walls. c) Conducting AFM image taken at room temperature in the same area and with a bias voltage applied to the sample of 2.75V (in the low current regime). d) Line scan across one domain wall in both the piezoresponse amplitude and the current images.}
\label{Figure1}
\end{figure}
Recently, the potential of domain walls has reached a new milestone, with the observation of conductivity at certain types of ferroelectric/ferroelastic domain walls in insulating multiferroic BiFeO$_3$\cite{Sei09,Sei10} thin films. Because domain walls in ferroelastic epitaxial thin films can be now controlled to great extent\cite{Jan09,Vlo07}, this discovery opens the door to novel nanometer-sized devices, as long as we are able to engineer and control the conduction paths in these materials. For that understanding of the origin of the observed conductivity is crucial. Although some work has already been done\cite{Sei10}, no clear picture exist. The present work provides significant progress in this direction.
\begin{figure}
\includegraphics[width=7cm]{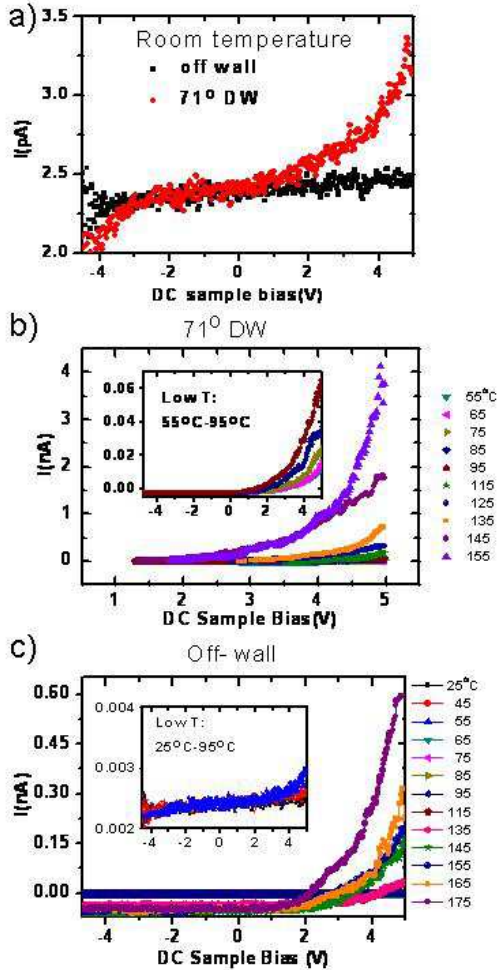}
\caption{(Color online)a) I-V curves at room temperature for 71$^o$ dws and off-wall (inside a domain). b)and c)show the I-V curves at different temperatures at 71$^o$ dw and off-wall, respectively. The lower temperature curves are plotted separately in the insets. }
\label{Figure2}
\end{figure}
Thin films of BiFeO$_3$ with thicknesses ranging between 40-70 nm have been slowly grown on SrRuO$_3$-buffered SrTiO$_3$ substrates, as described in ref.\cite{Dau10}. The as-grown samples contain monoclinic (pseudo-rhombohedral) crystallographic twin domains. From the eight possible twins\cite{Str98}, only the four variants pointing towards the SrRuO$_3$ electrode are present in the films, in agreement with other reports\cite{Jan09}. Most of the domain walls (dws) are 71$^o$ dws (where 71$^o$ is the approximate angular difference between the polarization vectors at both sides of the wall). In between two perpendicular sets of 71$^o$ dws, some 109$^o$ dws can also be found\cite{Dau10}. These ferroelectric/ferroelastic domains can be imaged by in-plane piezo-force microscopy (IP-PFM), using the metallic tip (Co-Cr coated Si) of an AFM microscope, as observed in Figure 1a)-b). The position and type of the dws can be accurately determined in this way. Simultaneously, the conduction through the sample, from the electrically grounded tip to the bottom SrRuO$_3$ electrode (held at variable bias voltages), can be locally measured and mapped using a TUNA$^{TM}$ amplifier (Bruker Corp.), which is designed to measure leakage currents from fA to $\mu $A. One such map is reproduced in figure 1c) and, when compared to the IP-PFM images, it reveals that the 71$^o$ (as well as the 109$^o$ domain walls) in these films are conducting at room temperature, while the domains are not conducting at this temperature. Figure 1d) shows both the in-plane piezoelectric response and the current when scanning across a 71$^o$ domain wall, showing the maximum conductivity in the middle of the wall (characterized by the minimum piezoelectric response). This is different from what is observed in other BiFeO$_3$ samples grown on DyScO$_3$, substrates, at higher deposition rates and with higher oxygen pressure by Seidel et al.\cite{Sei09}. In this case, conduction was found only at 109$^o$ dws. Conduction at 71$^o$ dws is advantageous because of their larger stability\cite{Ram11,Str98}.
\begin{figure*}
\includegraphics[width=12cm]{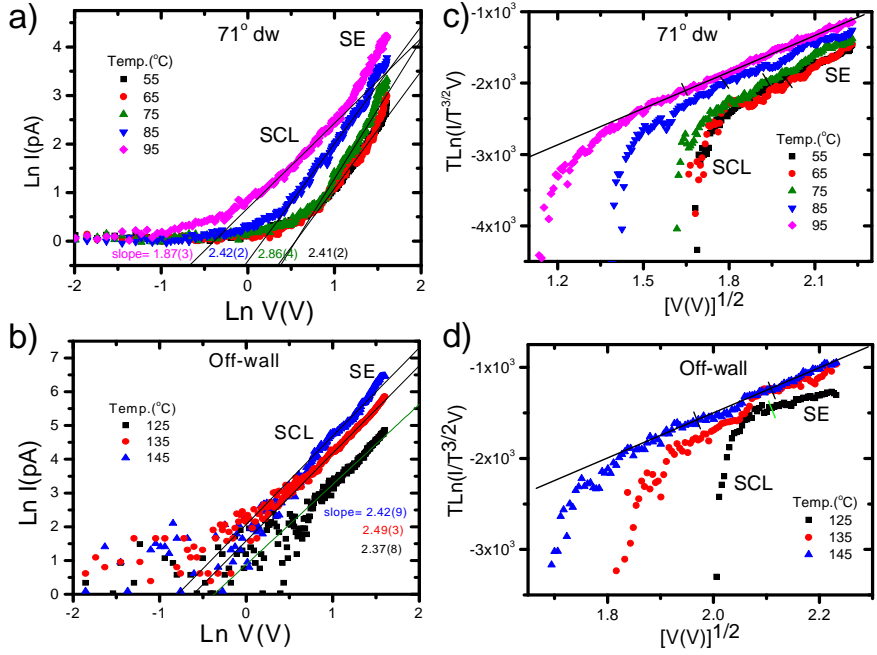}
\caption{(Color online). The I-V characteristics of figures 2b)and 2c) plotted in: (a,b) log-log scale in order to show linear behavior with slope ~2 in the space-charge limited (SCL) regime (the linear fits and the slope values are also shown)  and (c,d) as (TlnI/(T$^{3/2}$V) $versus $ V$^{1/2}$) to show linear behavior in the Schottky-Simmons emission regime\cite{Sim65}. In this regime, the slope, which should be common to all temperatures, is represented as a solid line. The short marks on the curves indicate the voltages from which the curves in a)-b) cannot be fitted in the SCL regime. The I-V curves are measured at the 71$^o$ dw (a, c) and off-wall (b,d). }
\label{Figure3}
\end{figure*}
Figure 2a) displays I-V curves for the 71$^o$ dws and off-wall at room temperature in linear scale, showing the rectifying diode behavior of the domain walls, as previously reported for the 109$^o$ walls\cite{Sei09}. The forward diode direction coincides with that of the electrical polarization and it indicates that carriers are n-type\cite{Blo94,Cho09}. In order to gain control on the currents, we investigate the conduction mechanisms by measuring the I-V curves at different temperatures (from 20 $^o$C to 200 $^o$C, in air). These curves are plotted in Figures 2b)-c) for 71$^o$ dws and off-wall, respectively. About 100 walls are investigated and all showed similar behavior. We have looked for signatures of interface-limited conduction, such as Fowler-Nordheim tunneling (ln(I/V$^2$) $\propto $E$^{-1}$) or Richardson-Schottky-Simmons emission (lnI/(T$^{3/2}$V) $\propto $ V$^{1/2}$)\cite{Sim65}, as well as of bulk-limited Poole-Frenkel emission (ln(I/V)$\propto $ V$^{1/2}$). The existence of current limitation due to built-in space charge (I$\propto $ V$^2$) is also investigated.

Interestingly, although appreciable conduction through the domains could only be observed at high temperatures (see figure 2), qualitatively similar behavior has been found for the two conduction paths (71$^o$ and off-wall) with two distinct regimes: For low voltages, conduction behaves like space-charge limited. This can be observed from the slope of the curves in figure 3a),b). The slope is around 2\cite{Note1} for voltages ranging from 1.5 V (where appreciable conduction starts) to about 3-3.5 V (depending of temperature). The temperature dependence in this regime shows Arrhenius thermal activation (see Figure 4). The activation energies extracted from the Arrhenius plots are consistent with defect states located 0.73(9)eV and 0.65(1)eV below the bottom of the conduction band, for 71$^o$ dws and off-wall, respectively. These values are too large for single electron traps at oxygen vacancies but they agree with those at clusters of oxygen vacancies in perovskites, and those reported for 109$^o$ domain walls\cite{Sei10}.

At large enough voltage (and temperatures), Schottky emission (SE) is observed\cite{Sim65}, as it can be elucidated from the linear behavior in Figure 3c)-d). The axes in the plots are chosen such as to display the same slope for the different temperatures. The slopes are shown with solid lines. It can be seen that the higher the temperature, the larger the voltage range for which SE is observed, as expected for termionic emission. The barrier heights obtained from fitting the Richardson-Schottky-Simmons equation\cite{Sim65}\cite{Note} differ considerably between the domain walls and the domains, being 0.8(1)eV for the 71$^o$ dws and 2.6(7)eV inside the domains. This large difference explains the lower currents measured at the domains and its reduction at room temperature below the sensitivity limit of the TUNA$^{TM}$ amplifier(5fA).
\begin{figure}
\includegraphics[width=7cm]{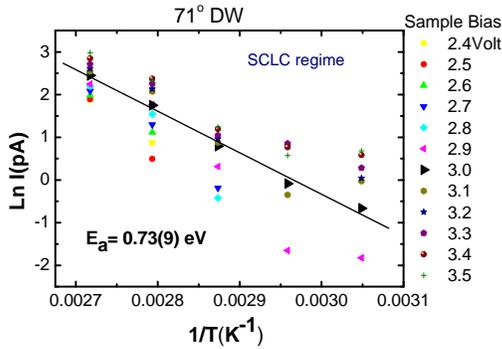}
\caption{(Color online)Arrehnius plot for the 71$^o$ domain walls in the low voltage regime, where the current shows a V$^2$ dependence (SCLC). An activation energy of 0.73(9)eV is obtained}
\label{Figure4}
\end{figure}
The Poole-Frenkel (PF) mechanism (field-assisted electron de-trapping) was also tested. The voltage dependence is similar to that of the SE, but the optical dielectric permittivity values obtained from the fits were considerably worse than those obtained from the SE equations. The dielectric permittivity at high frequencies in BiFeO$_3$ in 6.25\cite{Pin05}; a value of  6.5 $\pm $1.5 was obtained for 71$^o$ dws using the slope indicated in Figure 3c). With the PF fits, the obtained permittivity value is 0.4 $\pm $0.1. We observe then that the large currents (high voltage -high temperature) are interface-limited.

It is unusual that a space-charge-limited-like regime takes place at low currents. An explanation for this is provided by Blom et al.\cite{Blo94}: the ferroelectric surface charges themselves create a space-charge-like region giving rise to the I$\propto $ V$^2$ dependence. At large positive bias and higher temperatures, the positively charged oxygen vacancies would be able to diffuse towards the top surface to neutralize the polarization surface charge, decreasing the space charge limitation and finally allowing Schottky emission of electrons from the tip. Support for this model is provided by samples grown at a higher oxygen pressure (300 mbar instead of 100 mbar). In this case, the "space charge limited regime" (or more properly, the I$ \propto$ V$^2$ regime governed by the ferroelectric polarization) prevails up to the maximum measured bias (5V) for temperatures below 85$^o$ C and only above this temperature is an increase in the current observed. In this case, the behavior in the large temperature regime resembles that reported by Pintillie et al.\cite{Pin05}: Schottky emission with an unphysically small barrier of (0.14 eV, in our case) due to the influence of the ferroelectric polarization. This is indeed consistent with the fact that, in this sample with a lower number of oxygen vacancies present, the surface charges of the ferroelectric polarization are not completely neutralized by the oxygen vacancies.

The presence of n-type defect sates in the band gap (oxygen vacancies) will have two effects: first, to enable thermally activated electrons from defect state inside the band gap to be promoted to the conduction band, as discussed above, and second, the effect of reducing the Schottky barrier due to defect surface states\cite{Daw01}. This second effect is critical to control conduction at domains and domain walls, since it determines the large current regime. The role of oxygen vacancies is, thus, direct only in the low current regime. In the large current regime (the interesting one for applications) the oxygen vacancies do not directly provide the carriers, but they lower the Schottky barrier more in the domains walls (0.8(1) eV) than in the domains (2.6 eV), probably due an accumulation of oxygen vacancies in the domain walls. The carriers are electrons injected from the conductive AFM tip. A large number of vacancies must be present in the films in order to avoid the current to be limited by the ferroelectric polarization.
\begin{figure}
\includegraphics[width=5cm]{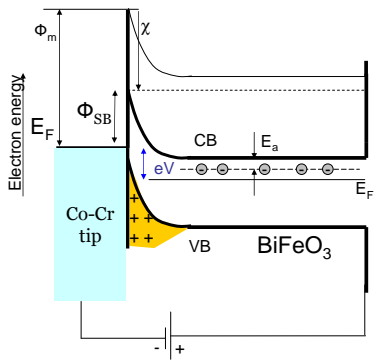}
\caption{(Color online)Sketch of the band bending at the tip-BiFeO$_3$ interface in the large voltage regime showing the Schottky barrier. In the low voltage regime, before emission takes places, electrons trapped in defect states are thermally promoted to the conduction band (activation energy E$_a$). In this low voltage regime, the ferroelectric polarization limits the current}
\label{Figure5}
\end{figure}
To conclude, we report that 71$^o$, naturally-formed, ferroelastic domain walls of BiFeO$_3$ grown on SrRuO$_3$-buffered SrTiO$_3$ provide well-defined conducting paths through the films, from the metallic tip acting as top electrode to the bottom SrRuO$_3$ electrode. Conduction is provided by n-type carriers and the mechanisms for conduction are the same both at the domain walls and at the domains: the low current regime is supported by thermally activated electrons from defect (oxygen vacancies) states and limited by the ferroelectric polarization; the large current regime is regulated by Schottky emission from the tip. The observed differences in conduction between the domains and the walls are due to the different barrier heights at the interface with the metallic tip. The large currents are therefore interface-limited and can be controlled by engineering the the work function of the injection electrode and the defect density in the BiFeO$_3$ layer.


\begin{references}
\bibitem{Note}In this work we are assuming that the BiFeO$_3$ is fully depleted and that the field at the interface (E$_o$) is proportional to the applied bias (V)\cite{Pin05,Zub06}.
\end{references}
\end{document}